\documentclass[prl,aps,twocolumn,tightenlines,notitlepage,nofootinbib,preprintnumbers,letterpaper,superscriptaddress]{revtex4-1} 
\pdfoutput=1
\usepackage{epsfig}
\usepackage{amsfonts}
\usepackage{amsmath}
\usepackage{graphicx}
\usepackage{color}
\usepackage{amssymb,amsmath,hyperref,slashed,color,graphicx,bm}
\usepackage[T1]{fontenc}
\usepackage{relsize}

\hypersetup{colorlinks,citecolor= nicegreen,linkcolor= nicered}
\definecolor{nicered}{rgb}{0.7,0.1,0.1}
\definecolor{nicegreen}{rgb}{0.1,0.5,0.1}

\def\Fermilab{Theoretical Physics Department, Fermilab, P.O. Box 500, Batavia, IL 60510, USA}
\def\ChicagoEFI{Enrico Fermi Institute, University of Chicago, Chicago, Illinois, 60637, USA}
\def\ChicagoKICP{Kavli Institute for Cosmological Physics, University of Chicago, Chicago, Illinois, 60637, USA}
\def\Barcelona{Institut de F\'isica d'Altes Energies (IFAE), The Barcelona Institute of Science and Technology (BIST),
Campus UAB, 08193 Bellaterra (Barcelona) Spain}
\def\Northwestern{Department of Physics and Astronomy, Northwestern University, Evanston, IL 60208, USA}
\def\ND{Department of Physics, University of Notre Dame, 225 Nieuwland Hall, Notre Dame, IN 46556, USA}

\begin{document}

\title{\Large
Electroweak Baryogenesis From Dark CP Violation
}
\author{Marcela Carena}
\affiliation{\Fermilab}
\affiliation{\ChicagoEFI}
\affiliation{\ChicagoKICP}

\author{Mariano Quir\'os}
\affiliation{\ND}
\affiliation{\Barcelona}

\author{Yue Zhang}
\affiliation{\Fermilab}
\affiliation{\Northwestern}

\date{\today}

\begin{abstract}
We present a novel mechanism of electroweak baryogenesis where \textit{CP} violation occurs in a dark sector, comprised of standard model gauge singlets, thereby evading the strong electric dipole moment constraints. In this framework, the background of time-like component of a new gauge boson $Z^\prime_\mu$, generated at electroweak temperatures, drives the electroweak sphaleron processes to create the required baryon asymmetry. We first discuss the crucial ingredients for this mechanism to work, and then show that all of them can be elegantly embedded in ultraviolet completions with spontaneously broken gauged lepton number. The models under consideration have a rich phenomenology and can be experimentally probed in leptophilic $Z^\prime$ searches, dark matter searches, heavy Majorana neutrino searches, as well as through hunting for new Higgs portal scalars in multi-lepton channels at colliders.
\end{abstract}

\preprint{FERMILAB-PUB-18-636-T}
\maketitle

The observed matter-antimatter asymmetry in the universe is widely believed to yield strong evidence for new phenomena beyond the standard model (SM) of particle physics. Electroweak baryogenesis (EWBG) is an elegant mechanism~\cite{Kuzmin:1985mm, Cohen:1990py, Cohen:1991iu, Farrar:1993sp, Quiros:1994dr,
Huet:1995mm, Huet:1995sh, Riotto:1995hh, Carena:1997gx, Quiros:1999jp,
Carena:2000id, Cline:2000nw, Carena:2002ss, Lee:2004we, Cline:2006ts} that generates the observed baryon asymmetry at the electroweak phase transition (EWPT).
This demands new physics close to the electroweak scale, to account for {\it CP} violating effects larger than those present in the SM.
Moreover, the requirement of a sufficiently strong EWPT, along with the precision measurements of the Higgs boson properties, demands
an extended scalar sector affecting the out-of-equilibrium processes.

The impressive recent progress in electric dipole moment (EDM) experiments~\cite{Andreev:2018ayy, Baron:2013eja,Griffith:2009zz,Baker:2006ts} imposes strong constraints on the required new sources of {\it CP} violation in SM extensions, such as two-Higgs-doublet models and supersymmetry~\cite{Shu:2013uua, Ipek:2013iba, Jung:2013hka, Abe:2013qla, Inoue:2014nva, Cheung:2014oaa, Bian:2014zka, Chen:2015gaa, Fuyuto:2015ida, Jiang:2015cwa, Blinov:2015sna, Balazs:2016yvi, Bian:2016zba, Chen:2017com, Cesarotti:2018huy, Egana-Ugrinovic:2018fpy}. This provides a strong motivation to consider {\it CP} violation triggered in dark sectors through SM gauge singlets~\cite{Cline:2017qpe}, which may naturally suppress contributions to EDMs. Such an enticing idea, however, leaves the challenging task of finding the suitable mechanism to transfer the {\it CP} violation from the dark sector to the visible sector, to successfully create the baryon asymmetry at electroweak temperatures.

In this {\it Letter}, we propose a new mechanism of EWBG where the transfer of {\it CP} violation to the visible sector is achieved by means of a vector boson $Z^\prime_\mu$ which couples to the SM leptons, and to dark fermions with {\it CP} violating Yukawa interactions involving additional SM singlet scalars. 
Such scalars may provide, through the Higgs portal, a sufficiently strong first-order EWPT.
The time-like component of the new gauge boson, $Z^\prime_0$, is {\it CP} odd and can transfer {\it CP} violation to the visible sector. During EWPT, the {\it CP} violating source yields a nonzero background $\langle Z^\prime_0\rangle$, which acts as a chemical potential for the SM leptons, providing a thermal equilibrium lepton number asymmetry. In the absence of any primordial asymmetries, such source term, through the electroweak sphaleron processes, generates an equal amount of baryon and lepton number asymmetries, which freeze in after the EWPT is completed and the sphalerons become inactive.
We point out that the current which couples to $Z^\prime_\mu$ must be anomalous with respect to the SM weak interaction during the EWBG epoch.

In the following, we will present the main ingredients of the proposed EWBG mechanism, and briefly explore its phenomenological consequences.
We refer to the companion paper~\cite{OurLongPaper} for a more detailed, quantitative presentation.

\medskip
\noindent{\it The Need of An Anomalous Current Coupled to $Z^\prime$.}\ \
The effective Lagrangian relevant to our discussion contains the couplings of a vector boson $Z^\prime_\mu$ to a current involving SM leptons and quarks, $J^\mu$,
\begin{eqnarray}
&&\mathcal{L}_{eff} = \mathcal{L}_{\rm SM} - \frac{1}{4} Z_{\mu\nu}^\prime Z^{\prime\mu\nu} + \frac{1}{2} M_{Z'}^2 Z_\mu^\prime Z^{\prime\mu} + g^\prime Z^\prime_\mu J^\mu \ , \label{Leff} \\ 
&&J^\mu = \sum_{i=1}^{3} \left[ q_{L_{L_i}} \bar L_{L_{i}} \gamma^\mu L_{L_i} + q_{e_{R_i}} \bar e_{R_i} \gamma^\mu e_{R_i} + q_{\nu_{R_i}} \bar \nu_{R_i} \gamma^\mu \nu_{R_i} \right. \nonumber \\
&&\ \ \ +\left. q_{Q_{L_i}} \bar Q_{L_i} \gamma^\mu Q_{L_i} + q_{u_{R_i}} \bar u_{R_i} \gamma^\mu u_{R_i} + q_{d_{R_i}} \bar d_{R_i} \gamma^\mu d_{R_i} \right],\label{eq:Current}
\end{eqnarray}
where $L_{L_{i}}, e_{R_i}$ and $\nu_{R_i}$ are the SM leptons and right-handed neutrinos,
$Q_{L_i}, u_{R_i}$ and $d_{R_i}$ are SM quarks, 
and $q_F$ is the charge of the corresponding fermion $F$.

We assume that the above effective Lagrangian describes a period of the early universe when EWBG occurs. Moreover,
we further assume that the vector field $Z^\prime_\mu$ develops a time-like background, $\langle Z^\prime_0\rangle\neq 0$, sourced by a $U(1)$ charge density, whose origin will be addressed later on.
Through Eq.~(\ref{Leff}) the $Z^\prime_0$ background acts as a chemical potential for the fermions. Of particular interest to us are the SM quark and lepton doublets. 
In the presence of a chemical potential, if the lepton or quark number were allowed to change independently, the particle-antiparticle number density asymmetry, defined as $\Delta n_F \equiv n_F- n_{\bar F}$ with $F=L_{L_{i}}, Q_{L_i}$, will be generated and evolve toward its thermal equilibrium value
\begin{equation}\label{ChemicalPotential}
\Delta n_{F}^{\rm EQ} = \frac{2 N_c}{3} T_c^2 g^\prime q_{F} \left\langle Z^\prime_0 \right\rangle \ ,
\end{equation}
where $N_c=3\ (1)$ for quarks (leptons) is the color factor, whereas $T_c$ is the EWPT critical temperature.
This expression is exact in the electroweak symmetric phase where all SM fermions are massless.

Within the context of EWBG, there is only one process where the lepton ($L$) and baryon ($B$) numbers are simultaneously violated -- the electroweak sphalerons. 
Each sphaleron process violates $B+L$ but conserves $B-L$ among the left-handed $SU(2)_L$ doublets, where the baryon and lepton asymmetries are defined as
\begin{equation}
\Delta n_{B_L} \equiv \frac{1}{3} \sum_{i=1}^{3} \Delta n_{Q_{L_i}}, \ \ \ \Delta n_{L_L} \equiv \sum_{i=1}^{3} \Delta n_{L_{L_i}} \ .
\end{equation}
The $(B\pm L)_L$ asymmetries satisfy the Boltzmann equations
\begin{eqnarray}\label{RateEquation}
\begin{split}
&\frac{\partial \Delta n_{(B+L)_L}}{\partial t}= \Gamma_{\rm sph} \left( \mathcal{S} -\Delta n_{(B+L)_L} \rule{0mm}{4mm}\right) \ , \\
&\frac{\partial \Delta n_{(B-L)_L}}{\partial t} =0, \ \ \ \ \ \mathcal{S} = \sum_{i=1}^3\left(\Delta n_{L_{L_i}}^{\rm EQ} + \Delta n_{Q_{L_i}}^{\rm EQ}\right) \ ,
\end{split}
\end{eqnarray}
where $\Gamma_{\rm sph} \simeq 120 \alpha_w^5 T_c$ is the sphaleron rate in the electroweak unbroken phase~\cite{Bodeker:1999gx} and it is exponentially suppressed in the broken phase.
$\mathcal{S}$ serves as the source for creating a net $B+L$ asymmetry, with $\Delta n_{L_{L_i}}^{\rm EQ}$ and $\Delta n_{Q_{L_i}}^{\rm EQ}$ contributing to it democratically, the same way as sphalerons act on every $SU(2)_L$ doublet.

Starting from a primordially symmetric universe implies, 
%
$\Delta n_{B_L} = \Delta n_{L_L} = (1/2) \Delta n_{(B+L)_L} $.
%
With this, the first equation in (\ref{RateEquation}) simplifies to
\begin{eqnarray}\label{RateEquation2}
\frac{\partial \Delta n_{B_L}}{\partial t} = \Gamma_{\rm sph} \left( \frac{1}{2} \mathcal{S} - \Delta n_{B_L}
\rule{0mm}{4mm}\right) \ ,
\end{eqnarray}
and from Eq.~(\ref{ChemicalPotential}) it is straightforward to derive
\begin{eqnarray}\label{ThermalnEQ}
\mathcal{S} = \frac{2}{3} T^2 g^\prime \, \sum_{i=1}^3 \left( q_{L_{L_i}} + 3 q_{Q_{L_i}}  \rule{0mm}{4mm}\right)  \left\langle Z^\prime_0 \right\rangle\, .
\end{eqnarray}

Remarkably Eq.~(\ref{ThermalnEQ}) is proportional to the 
non-conservation of the current $J^\mu$, {\it i.e.}, the coefficient appearing its
chiral anomaly with respect to $SU(2)_L^2$,
\begin{eqnarray}
\partial_\mu J^{\mu} \propto \sum_{i=1}^3 \left( q_{L_{L_i}} + 3 q_{Q_{L_i}}\right) {\rm tr}(W\widetilde W) \ ,
\end{eqnarray}
where $W$ ($\widetilde W$) is the $SU(2)_L$ field (dual) strength.
Hence we have found a \textit{necessary} condition for the proposed EWBG mechanism to work, namely, the current to which the $Z^\prime_\mu$ couples must be anomalous with respect to $SU(2)_L^2$.
Had the charges $q_F$ in Eq.~(\ref{eq:Current}) been arranged such that the current $J^\mu$ were conserved,
the source term in Eq.~(\ref{RateEquation2}) would have vanished and, 
in turn, no net baryon asymmetry would have been created. 

The effective Lagrangian of Eq.~(\ref{Leff}) can be obtained from a UV complete $U(1)$ gauge theory whose gauge boson is $Z^\prime_\mu$ and 
$q_F$ are the corresponding SM fermion $U(1)$ charges. 
In the case of a non-conserved $J^\mu$, additional fermions (anomalons) are required to render the $U(1)$ anomaly free.
The total current of the $U(1)$ gauge symmetry is the sum of $J^\mu$ in Eq.~(\ref{eq:Current}) and that of the anomalons, $J^\mu_a$, such that the anomaly cancellation
condition imposes, $\partial_\mu (J^\mu + J^\mu_a)=0$.
The anomalon fields, once introduced, will also contribute to the source term $\mathcal{S}$, Eq.~(\ref{ThermalnEQ}).
Here, however, we assume that the $U(1)$ gauge symmetry is spontaneously broken above the electroweak scale ({\it e.g.}, at TeV scales), and that the 
anomalons get symmetry breaking masses. If the anomalons have masses much larger than the EWPT temperature, their population, as well as their impact on the electroweak sphalerons, will become Boltzmann suppressed.
In such case, the anomalons, although canceling the gauge anomalies, have negligible contribution to $\mathcal{S}$.

Good candidates for such a $U(1)$ symmetry include gauged lepton number, baryon number, or any flavor dependent combination of the two that remains anomalous, within the SM, with respect to $SU(2)_L^2$.
In contrast, the proposed EWBG mechanism cannot work if the $U(1)$ is already anomaly free given the SM fermion content (plus right-handed neutrinos),
for example, $B-L$, $L_\mu-L_\tau$, etc.
In the following we discuss the realization of our EWBG mechanism in a UV complete model with gauged lepton number, $U(1)_\ell$.
As we shall see, a dark matter candidate also naturally emerges in the theory.

As another useful remark, interestly the above discussion remains valid even if the $\langle Z^\prime_0\rangle$ background is space-time inhomogeneous. 
Indeed, we will consider a first order EWPT which temporarily creates $\langle Z^\prime_0\rangle$ in front of the expanding bubble walls.

\medskip
\noindent{\it CP Violation and the Electroweak Phase Transition.} \ \ 
We will now address the origin of the $Z^\prime_0$ field background, as well as the dynamics of the EWPT.
In analogy to a static electric potential, the $\langle Z^\prime_0\rangle$ background is \textit{C}, {\it CP} and {\it CPT} odd, and can be generated by a net $U(1)_\ell$ charge distribution near the bubble wall. To this end we introduce a fermionic particle 
$\chi$ with {\it CP} violating microscopic interactions with the bubble wall. 
Since $\chi$ is a SM gauge singlet that cannot couple to the Higgs field
through renormalizable interactions, we will introduce a SM scalar singlet $S$ to interact with it,
\begin{eqnarray}\label{darkYukawa}
\bar \chi_L ( m_0 + \lambda e^{i\theta_\lambda} S ) \chi_R + {\rm h.c.}
\end{eqnarray}

Within the bubble wall, the Higgs VEV turns on, while the $S$ VEV simultaneously turns off.
Such a transition to the electroweak broken phase has been studied and involves a two-step process from the original vacuum with $\langle S\rangle=\langle H\rangle=0$~\cite{Espinosa:2011ax, Patel:2013zla, Cheung:2013dca, Curtin:2014jma, Cline:2017qpe}.
The necessary ingredient to allow for a strongly first order EWPT is a sizable scalar quartic term, $|S|^2|H|^2$, by which the Higgs becomes a portal to the dark sector.

We will consider the following ansatz for the $S$ profile across the bubble wall,
$|S(z)| = s_0 \left[1 + \tanh (z/L_\omega) \rule{0mm}{3mm}\right]/2$.
The coordinate $z$ is defined in the rest frame of the bubble wall
which is located at $z=0$, whereas $L_\omega$ is the wall width.
Observe that to accommodate a physical {\it CP} violating effect through Eq.~(\ref{darkYukawa}), we need a scalar potential that fixes the phase of $S$.
During the EWPT, the VEV of $S$ contributes to the $\chi$ mass through Eq.~(\ref{darkYukawa}), whereas the bare mass term $m_0$ has its origin in the spontaneous breaking of $U(1)_\ell$.
If the two mass terms carry different, space-time dependent phases, the dispersion relations of $\chi_L, \chi_R$ and their antiparticles will be modified in a {\it CP} violating way.
This affects the phase space distributions of such particles and yields a non-trivial solution to the corresponding diffusion equations,
leading to net number density asymmetries in $\chi_L, \chi_R$,
\begin{eqnarray}
\Delta n_\chi (z) \equiv n_{\chi_L} - n_{\chi^c_L} = - (n_{\chi_R}-n_{\chi^c_R}) \neq 0 \ .
\end{eqnarray}
The spatial distribution of $\Delta n_\chi(z)$ will peak around the bubble wall.
For details on solving the diffusion equations and the numerical computation of $\Delta n_\chi (z)$,
we refer the reader to the companion paper~\cite{OurLongPaper}.
If $\chi_L$ and $\chi_R$ carry different $U(1)_\ell$ quantum numbers, the above chiral asymmetries will give a net $U(1)_\ell$ charge density distribution around the bubble wall,
\begin{equation}\label{NetELLCharge Density}
\rho_\ell (z) = (q_{\chi_L}-q_{\chi_R}) \Delta n_\chi (z) \ .
\end{equation}

Neglecting the curvature of the bubble wall, the $\langle Z^\prime_0\rangle$ background sourced by $\rho_\ell$ can be calculated in cylindrical coordinates to be
\begin{equation}\label{eq:Zbackground}
\langle Z'_0(z)\rangle = \frac{g' }{2M_{Z'}} \int_{-\infty}^\infty d y\ \rho_\ell(y)\ e^{-M_{Z'} |z-y|}\ .
\end{equation}
Given this $\langle Z^\prime_0\rangle$ background, the final baryon asymmetry generated can be obtained by solving Eq.~(\ref{RateEquation2}),
\begin{equation}\label{eq:LeptonAsymmetry}
\Delta n_{B} =\frac{\Gamma_{\rm sph}}{v_\omega} \int^{\infty}_0 dz 
\ \mathcal{S}(z) \
e^{-\Gamma_{\rm sph} z/v_\omega} \ ,
\end{equation} 
where $v_\omega$ is the bubble wall expansion velocity. The parametric dependence in today's baryon to entropy ratio is,
$\eta_B = {\Delta n_{B}}/{s} \sim g'^2 N_g^2 T_c^3 L_\omega \alpha_W^5/(M_{Z'}^2 v_\omega)$.

\medskip
\noindent{\it UV Complete Models.} \ \
Next, we discuss unifying all the above ingredients for EWBG into a UV complete framework with gauged (anomaly free) lepton number symmetry, $U(1)_\ell$. There are several choices to define the lepton number $\ell$.
The most obvious one is to gauge all the three SM families universally by taking $\ell=L_e+L_\mu+L_\tau$. Alternatively, one could also gauge only two lepton flavors such as
$\ell=L_\mu+L_\tau$. 
We will consider these two cases as benchmark models.
The minimal set of new fermion content is given in Table~\ref{table1}~\cite{FileviezPerez:2010gw, Schwaller:2013hqa},
where $\mathtt{q}$ is an arbitrary real number. The index $i$ runs through $e,\mu,\tau$ ($\mu,\tau$) in the first
(second) model, and $N_g=3\ (2)$ defines the number of families charged under the $U(1)_\ell$, correspondingly.
\begin{table}[h]
\centering
\begin{tabular}{||c|c|c|c|c||}
\hline\hline
Particle & $SU(3)_c$ & $SU(2)_L$ & $U(1)_Y$ & $U(1)_\ell$ \\
\hline
$L_{L_i}$ & 1 & 2 & -1/2 & 1 \\
$e_{R_i}$ & 1 & 1 & -1 & 1 \\
$\nu_{R_i}$ & 1 & 1 & 0 & 1 \\
$L^\prime=(\nu_L', e_L')^T$ & 1 & 2 & -1/2 & $\mathtt{q}$ \\
$e^\prime_R$ & 1 & 1 & -1 & $\mathtt{q}$ \\
$\chi_R$ & 1 & 1 & 0 & $\mathtt{q}$ \\
$R^\prime=(\nu_R^{\prime\prime}, e_R^{\prime\prime})^T$ & 1 & 2 & -1/2 & $\mathtt{q}+N_g$ \\
$e^{\prime\prime}_L$ & 1 & 1 & -1 & $\mathtt{q}+N_g$ \\
$\chi_L$ & 1 & 1 & 0 & $\mathtt{q}+N_g$ \\
\hline
$\Phi$, $S$ & 1 & 1 & 0 & $N_g$ \\
\hline\hline
\end{tabular}
\caption{\it UV completion of the effective theory.}
\label{table1}
\end{table}

\begin{figure*}
\includegraphics[width=1\textwidth]{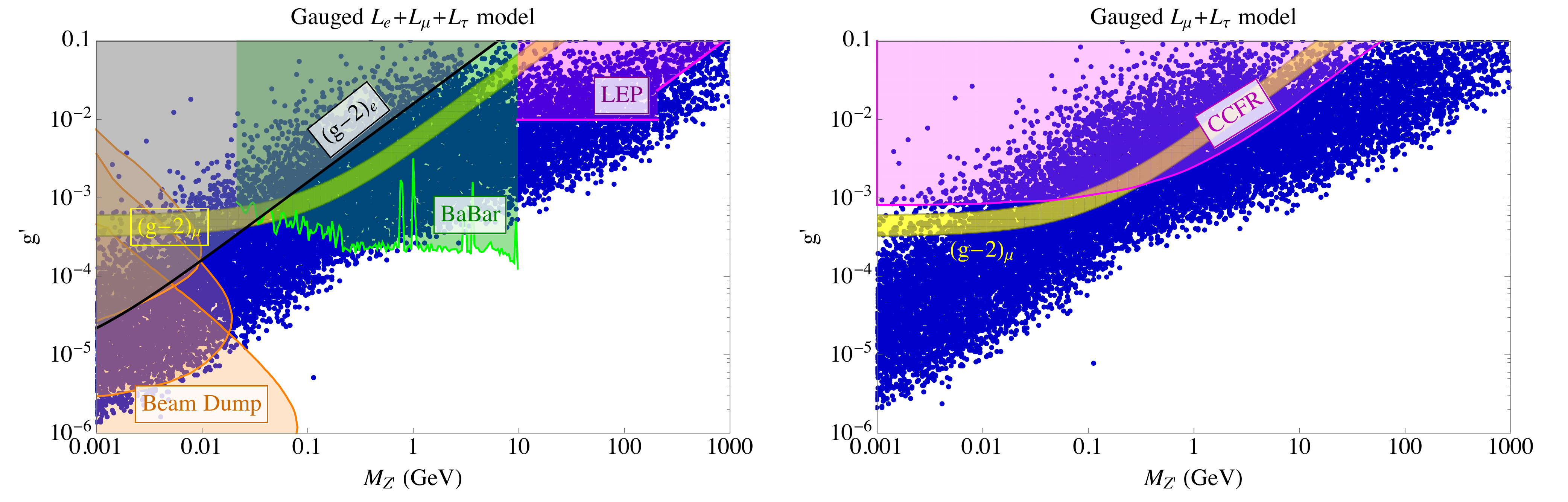}
\caption{\it 
We scan broadly over the model parameters to find points that allow for
successful EWBG as proposed in this work (blue points). We show various experimental constraints (LEP, BaBar, beam dump, electron and muon $g-2$, CCFR) by the correspondingly labeled shaded regions, in the gauged $L_e+L_\mu+L_\tau$ (left) and $L_\mu+L_\tau$ (right) models.}
\label{fig:moneyplots}
\end{figure*}

To spontaneously break the $U(1)_\ell$, and at the same time give masses to the new fermions, we introduce the complex scalar $\Phi$ carrying $U(1)_\ell$ number $N_g$.
We assume that $\Phi$ picks up a VEV, $v_\Phi/\sqrt2$, above the electroweak scale. This VEV gives mass to the gauge boson $Z^\prime$, $M_{Z^\prime} = N_g g^\prime v_\Phi/2$, which can still be light if the gauge coupling $g^\prime$ is sufficiently small.
We can also write down Yukawa couplings of the form,
\begin{equation}\label{PhiYukawa}
\left( c_L \bar R' L'  + c_e \bar e^\prime_L e^{\prime\prime}_R + c_\chi \bar \chi_L \chi_R \rule{0mm}{3.5mm}\right) \Phi + {\rm h.c.} \ ,
\end{equation}
which will give vectorlike masses (with respect to the SM) to the new fermions.
We assume $c_L \sim c_e$ to be large enough so that $L^\prime, R^\prime, e^{\prime\prime}_L, e^\prime_R$ are sufficiently heavy in comparison with the critical temperature of the EWPT. 
As noted earlier, the fermions $L^\prime$ and $R^\prime$ are needed to cancel the $U(1)_\ell\otimes SU(2)_L^2$ gauge anomaly, whereas decoupling them from the thermal bath provides the necessary condition for our EWBG mechanism to work.
On the other hand, we assume the parameter $c_\chi$ to be sufficiently small so that $\chi$ is light and remains populated in the thermal bath during the EWPT.
Their $U(1)_\ell$ charges are fixed in the UV theory: $q_{\chi_L}=\mathtt{q}+N_g$, $q_{\chi_R}=\mathtt{q}$, and their difference does not depend on $\mathtt{q}$.
This helps to eliminate a free parameter from Eq.~(\ref{NetELLCharge Density}), so that $\rho_\ell (z) = N_g \Delta n_\chi$.

The fermion $\chi$ will source {\it CP} violation when it interacts with the expanding bubble wall. 
The $\chi$-$\Phi$ interaction in Eq.~(\ref{PhiYukawa}) is responsible for generating the $m_0=c_\chi v_\Phi/\sqrt2$ mass term in Eq.~(\ref{darkYukawa}).
In addition, as discussed before, another complex scalar $S$ with the same quantum numbers as $\Phi$ and a scalar potential that fixes its phase is required
to yield a physical {\it CP} phase, barring the redefinitions of fermion fields.
Moreover, such a complex scalar will also be responsible for the strong EWPT through a two-step phase transition. 

To summarize, we have argued that all the key ingredients for our EWBG mechanism can be elegantly accommodated in a UV theory of gauged lepton number.  The UV completion is useful in shedding light into the parameters of the low energy effective theory.

\medskip
\noindent{\it Neutrino Cosmology.} \ \
It is worth commenting on the neutrino sector and implications of cosmological measurements on additional neutrino degrees of freedom, $\Delta N_{eff}$~\cite{Barger:2003zh, Ghosh:2010hy, Barger:2010iv, Hamann:2011ge}, for the two benchmark models.
The $U(1)_\ell$ gauge interaction could thermalize, in the very early universe, all the new fermions charged under it, and in particular the right-handed neutrinos $\nu_{R_i}$. 
To avoid an excessive contribution to $\Delta N_{eff}$, one option is to make the $U(1)_\ell$ interaction decouple early enough, preferably above the QCD phase transition temperature, $T_{\rm QCD} \sim 100\,$MeV. This implies $v_\Phi \gtrsim 10\,$TeV if $M_{Z^\prime}\gg T_{\rm QCD}$, or $g^\prime\lesssim10^{-5}$ if $M_{Z^\prime}\ll T_{\rm QCD}$. The other option is to implement the seesaw mechanism by giving Majorana masses to $\nu_{R_i}$. 
If all the $\nu_{R_i}$ are heavier than $\sim 500\,$MeV, they will decay before the big-bang nucleosynthesis and have no effect in $\Delta N_{eff}$~\cite{Berryman:2017twh}. 
Interestingly, this option could easily be achieved in the gauged $L_\mu+L_\tau$ model where both $\Phi$ and $S$ have charge $N_g=2$.
The experimental search for heavy Majorana neutrinos is of great phenomenological interests~\cite{Atre:2009rg}, especially as the new $U(1)_\ell$ gauge interaction allows them to be more copiously produced. We will investigate this exciting opportunity in a future work.

\medskip
\noindent{\it Experimental Probes.} \ \
Here we summarize the phenomenological predictions unique to our EWBG mechanism and discuss the present experimental bounds in the two benchmark models presented above. 

-- The main motivation for this work is to provide the necessary amount of \textit{CP} violation for baryogenesis, without being in tension with EDM measurements. In the gauged $U(1)_\ell$ UV complete models presented here, the fermion field $\chi$ responsible for {\it CP} violation is a SM singlet, thereby eliminating its Barr-Zee type~\cite{Barr:1990vd} contribution to EDMs. In Ref.~\cite{OurLongPaper}, we will show that the leading contribution arises at the four-loop level.

-- In our EWBG mechanism, the $Z^\prime_\mu$ gauge boson transmits the dark CP violation to the SM sector, and thereby the generation of the observed baryon asymmetry, $\eta_B \simeq 0.9\times10^{-10}$, restricts the values of $M_{Z'}$ as a function of its gauge coupling $g'$. In Fig.~\ref{fig:moneyplots}, the blue points are obtained from a scan over the parameter space that can account for the correct $\eta_B$. The allowed $Z'$ masses are in the MeV to TeV range, with decreasing $g'$ values for lighter $Z'$, in agreement with the parametric dependence estimated below Eq.~(\ref{eq:LeptonAsymmetry}). In the left (right) panel of Fig.~\ref{fig:moneyplots}, we show the gauged $L_e+L_\mu+L_\tau$ ($L_\mu+L_\tau$) benchmark models.

-- The search for $Z^\prime$ provides a handle on these EWBG scenarios.
In the gauged $L_e+L_\mu+L_\tau$ model, the $Z^\prime$ has a coupling to the electron, which is subject to constraints from electron $g-2$, $e^+e^-$ colliders (LEP, BaBar) and electron beam dump experiments~\cite{Alexander:2016aln}, as shown by the correspondingly labeled shaded regions in the left panel of Fig.~\ref{fig:moneyplots}.
On the other hand, the gauged $L_\mu+L_\tau$ model is free from the above constraints. There is, however, a relevant constraint from neutrino trident production~\cite{Altmannshofer:2014pba} which excludes the magenta region (labeled by CCFR) in the right panel of Fig.~\ref{fig:moneyplots}. Interestingly, Fig.~\ref{fig:moneyplots} shows that there is a region of parameter space that can explain the muon $g-2$ anomaly (yellow band), and in the case of the $L_\mu+L_\tau$ model, such a region is allowed and favored by EWBG, with $M_{Z^\prime}\lesssim 200\,$MeV.

-- The models considered here provide a dark matter candidate, $\chi$. The $U(1)_\ell$ gauge invariance implies that the SM singlet fermion $\chi$ can only interact with the SM particles via the $Z^\prime$ exchange, making $\chi$ a leptophilic dark matter candidate~\cite{Fox:2011fx}.
Its thermal relic density and detection prospects will be discussed in detail in Ref.~\cite{OurLongPaper}. 

-- It is possible to search for the dark scalar $S$ at high energy colliders, where it can be pair produced through the Higgs portal interaction.
If $S$ is lighter than twice the $\chi$ mass, it must decay via a $\chi$ loop into a pair of $Z^\prime$ bosons, yielding four leptons in the final state.
The $S$-$\chi$ interaction is inherently {\it CP} violating and such decay can provide a test of dark {\it CP} violation via interference effects in the golden $4\ell$-channel.

\medskip
\noindent{\it Summary.}\ \
We have proposed a novel mechanism for EWBG in which the {\it CP} violation occurs in a dark sector and is transmitted to the observable sector via 
the time-like background of a $Z^\prime_\mu$ vector boson during a strong first-order EWPT. The $Z^\prime_\mu$ is the gauge boson of a $U(1)_\ell$ gauge symmetry, and couples to an anomalous SM lepton number current. After the spontaneous $U(1)_\ell$ symmetry breaking, the new $SU(2)_L$ doublet fermions required to render the theory anomaly free become massive and decouple from the thermal bath before the EWPT. Because the {\it CP} violating interactions are active in the dark sector, its effects on EDMs are highly suppressed and evade present bounds. We show two benchmark scenarios with gauged $U(1)_{L_e+L_\mu+L_\tau}$ and $U(1)_{L_\mu+L_\tau}$ symmetries which provide concrete examples of UV completions. 
The models under consideration provide a rich phenomenology that can be probed in searches for leptophilic $Z^\prime$, dark matter, heavy Majorana neutrinos, and new scalars in multi-lepton channels at the LHC or prospective high energy colliders.

\medskip
\begin{acknowledgements}
We thank Zackaria Chacko, James Cline, Bogdan Dobrescu, Paddy Fox, Ian Low, David Morrissey, Tim Tait and Carlos Wagner for useful discussions and comments.
This manuscript has been authored by Fermi Research Alliance, LLC under Contract No. DE-AC02-07CH11359 with the U.S. Department of Energy, Office of Science, Office of High Energy Physics. 
Work at University of Chicago is supported in part by U.S. Department of Energy grant number DE-FG02-13ER41958. The work of M.Q. is partly supported by Spanish MINEICO
under Grant CICYT-FEDER-FPA2014-55613-P and FPA2017-88915-P, and by the Severo Ochoa Excellence Program of MINEICO under Grant SEV-2016-0588.   
The work of Y.Z. is also supported by the DoE under contract number DE-SC0007859.
Y.Z. would like to thank the Aspen Center for Physics, which is supported by National Science Foundation grant PHY-1607611, where part of this work was performed, and Colegio De Fisica Fundamental E Interdisciplinaria De Las Americas (COFI) for a travel support during the completion of this work.
\end{acknowledgements}

\bibliography{References}

\end{document}